\documentclass[fleqn,usenatbib,useAMS]{mnras}

\usepackage{savesym}
\usepackage{amsmath}
\savesymbol{iint}
\usepackage{txfonts}
\restoresymbol{TXF}{iint}

\usepackage{amssymb}
\usepackage[T1]{fontenc}
\usepackage{natbib}
\usepackage{hyperref}
\usepackage{epsfig}
\usepackage{graphicx}
\usepackage{tablefootnote}
\usepackage{caption}

\newcommand{\mc}{\mathcal}
\newcommand{\be}{\begin{equation}}
\newcommand{\ee}{\end{equation}}
\newcommand{\bea}{\begin{eqnarray}}
\newcommand{\eea}{\end{eqnarray}}
\newcommand{\nn}{\nonumber}

\title[Improving Statistical Sensitivity of X-ray Searches for Axion-Like Particles]{Improving Statistical Sensitivity of X-ray Searches for Axion-Like Particles}
\author[J. Conlon and M. Rummel]{
Joseph P. Conlon$^{1}$\thanks{E-mail: joseph.conlon@physics.ox.ac.uk}
and Markus Rummel$^{2, 3}$\thanks{E-mail: rummelm@mcmaster.ca}
\\
$^{1}$Rudolf Peierls Centre for Theoretical Physics, 1 Keble Road, Oxford, OX1 3NP, UK\\
$^{2}$Department of Physics and Astronomy, McMaster University, Hamilton, Ontario, L8S 4M1, Canada\\
$^{3}$Perimeter Institute for Theoretical Physics, Waterloo, Ontario, N2L 2Y5, Canada
}
\date{\today}

\pubyear{2017}

\begin{document}
\label{firstpage}
\pagerange{\pageref{firstpage}--\pageref{lastpage}}
\maketitle

\begin{abstract}
X-ray observations of bright AGNs in or behind galaxy clusters offer unique capabilities to constrain axion-like particles.
Existing analysis technique rely on measurements of the global goodness-of-fit.
We develop a new analysis methodology that improves the statistical sensitivity to ALP-photon oscillations by isolating
the characteristic quasi-sinusoidal modulations induced by ALPs. This involves analysing residuals in wavelength 
space allowing the Fourier structure to be made manifest as well as a machine learning approach. For telescopes with microcalorimeter resolution, simulations suggest these methods give an additional factor of two in sensitivity to ALPs compared to previous approaches.
\end{abstract}


\section{Introduction}
\label{introduction}

Axion-like particles are one of the simplest and best motivated extensions of the Standard Model.
They arise as a generalisation of the QCD axion, which was introduced as an appealing solution to the strong CP problem of QCD (\cite{PecceiQuinn, Weinberg, Wilczek}) - a recent review of axion physics is \cite{Marsh}.
While the original QCD axion requires a coupling to the strong force, it is also interesting to consider more general \emph{axion-like particles} (ALPs) that couple only to electromagnetism. Such ALPs arise generally in string compactifications (for example, see \cite{Conlon:2006tq, Svrcek:2006yi, Arvanitaki:2009fg, Cicoli:2012sz}). An ALP $a$ interacts with photons via the coupling:
\bea
&  g_{a\gamma\gamma}\, a\, \vec{E}\cdot\vec{B}, &
\eea
where $g_{a\gamma\gamma}$ is a dimensionful coupling parametrizing the strength of the interaction and $\vec{E},\vec{B}$ are the electric and magnetic fields, respectively. While we refer in this paper to ALPs, the physics is also relevant for photophilic models of the QCD axion in which the mass is much smaller (or the photon coupling significantly enhanced) compared to naive expectations, such as those in \cite{Farina, Agrawal}.

As they attain masses only by non-perturbative effects, ALPs  naturally have extremely small masses. The relevant physics is described by the Lagrangian
\bea
\mathcal{L} & = & \frac{1}{2} \partial_{\mu} a\, \partial^{\mu} a + \frac{1}{2} m_a^2 a^2 + g_{a\gamma\gamma}\,a\,\vec{E}\cdot\vec{B}.
\eea
The ALP-photon coupling produced by the $a\vec{E}\cdot\vec{B}$ interaction implies that, within background magnetic fields, the ALP state $a$ has a 2-particle interaction with the photon $\gamma$. Under this mixing, the `mass' eigenstates of the Hamiltonian are a mixture of the photon and ALP `flavour' eigenstates, causing oscillation between the modes in a way analogous to neutrino oscillations (\cite{Sikivie:1983ip, RS}).

ALP-photon conversion is enhanced by large magnetic field coherence lengths. As it extends over megaparsec scales and contains coherence lengths up to tens of kiloparsecs, this makes the intracluster medium of
galaxy clusters particularly efficient (\cite{Burrage:2009mj, Angus:2013sua, Conlon:2013txa, Powell:2014mda, Schlederer:2015jwa, Conlon:2015uwa}). This has allowed the use of X-ray point sources, located in or behind clusters, to produce leading bounds on ALP parameter space (\cite{Wouters:2013hua, 160501043, 1704.05256, Marsh:2017yvc, Conlon:2017ofb, Chen:2017mjf}). While X-rays are particularly good at probing ALPs with $m \lesssim 10^{-11} {\rm eV}$, more massive ALPs can be probed via gamma-ray astronomy - see \cite{Wouters:2012qd, Ajello, Payez, MajumdarHorns} for some related work in different wavebands.

The aim of this work is to improve the theoretical tools available for the analysis of ALP-photon conversion, and in particular the statistical methodology used to analyse data. Existing approaches to bounding ALP-photon conversion involve, roughly, a comparison of the overall goodness-of-fit (as measured by the reduced $\chi^2$) of models with ALPs compared to models without ALPs. For sufficiently large ALP-photon coupling, models with ALPs lead to a bad fit.

However, this approach makes little use of the actual structure of ALP-photon conversion. While the precise form of ALP-photon conversion depends on the (unknown) magnetic field along the line of sight, there is a quasi-sinusoidal oscillatory structure that is common to all realisations of the magnetic field. The aim of this paper is to develop analysis techniques to isolate this structure, and thereby allow
greater sensitivity to constraining ALPs. This development of improved statistical methods is also important in view of the
launch over the next decade of microcalorimeter-based X-ray telescopes such as XARM and ATHENA; the use of an
overall $\chi^2$ fit would then fail to take full advantage of the energy resolution provided by such instruments.

With an eye on the opportunities that will come from future satellites such as XARM and ATHENA, we
focus on energies and parameter ranges that are applicable for the case of
X-ray photons converting to ALPs within galaxy cluster magnetic fields.

\section{Statistics of ALP-Photon Conversion}
\label{alps}

To develop the analysis formalism, we start with the theory of photon-ALP conversion for a single magnetic field domain.
For an X-ray photon passing through a single magnetic field domain (\cite{Sikivie:1983ip, RS}), 
\bea
P(\gamma \to a) & = & \frac{\Theta^2}{1 + \Theta^2} \sin^2 \left( \Delta \sqrt{1 + \Theta^2} \right),
\eea
where
\bea
\Theta & = &  \frac{2 g_{a\gamma\gamma} B_{\perp} \omega}{\left( \omega_{pl}^2 - m_a^2 \right) }, \\
\delta & = & \frac{ \left( m_a^2 - \omega_{pl}^2 \right) L}{4 \omega}.
\eea
The numerical values of $\Theta$ and $\Delta$ are
\bea
\Theta & = & 0.02 \left( \frac{B}{1 \mu G} \right) \left( \frac{\omega}{1 {\rm keV}} \right)
\left( \frac{10^{-3} {\rm cm}^{-3}}{n_e} \right) \left( \frac{g_{a \gamma \gamma}}{10^{-12} {\rm GeV}^{-1}} \right), \\
\Delta & = & 0.54 \left( \frac{n_e}{10^{-3} {\rm cm}^{-3} } \right) \left( \frac{L}{10 {\rm kpc}} \right) \left( \frac{1 {\rm keV}}{\omega} \right).
\eea
As values of $g_{a \gamma \gamma}$ significantly greater than $10^{-12} {\rm GeV}^{-1}$ are already excluded (\cite{160501043, Marsh:2017yvc, 1704.05256}),
this implies that for typical parameters at X-ray energies $\Theta^2 \ll 1$ (for regions with $B \sim 25 - 50 \mu{\rm G}$ such as the centre of cool core clusters, $n_e \gtrsim 10^{-2} {\rm cm}^{-3}$).
In this limit we can simplify the single-domain conversion probability to
\bea
P(a \to \gamma) & = & \Theta^2 \sin^2 \Delta + \mc{O}(\Theta^4),
\eea
which as a function of energy behaves as
\bea
P(\omega) & \sim & \epsilon \left( \frac{\omega}{\omega_0} \right)^2 \sin^2 \left[ \Delta_0 \left( \frac{\omega_0}{\omega} \right) \right] + \mc{O}(\Theta^4),
\eea
where $\epsilon$ and $\Delta_0$ are constants, and $\omega_0$ is a reference energy which we take to be 1 keV.

Along a sightline through a cluster, the magnetic field undergoes many reversals.
We model the cluster magnetic field as a sequential series of domains, each differing in terms of field strength, direction and coherence length.
As the properties of each domain are independent of the previous one,
we can obtain the total survival probability by multiplying the survival probability for each domain.\footnote{Although the axion propagation equations evolve amplitude not probability, as different domains add incoherently it is sufficient to combine probabilities across domains.}

We note that this model consists of separate sequential magnetic field domains, each with a different coherence length and magnetic field. A more accurate model would involve, instead of distinct domains, many different scales simultaneously present in the magnetic field.
However, simulations of photon-ALP conversion in such cases do not show major qualitative differences from the case where the different scales are represented by separate sequential domains (\cite{Angus:2013sua}). As it is conceptually and calculationally much simpler, and does not appear to be too misleading, we shall therefore work with a magnetic field model of separate and sequential domains.

In this paper we will always assume that the overall photon-axion conversion probability is very small (less than ten percent).
This is not a significant restriction - if it is not the case, then the more advanced techniques
of this paper are unnecessary as
a simple test such as an overall $\chi^2$ fit will already show that the standard astrophysical fit fails to describe the data.

The combination of the $\gtrsim \mc{O}(100)$ domains required to describe a Mpc-scale cluster and an overall
conversion probability of less than 10\% implies that
the conversion probability within any individual single domain
is highly suppressed. This allows the overall photon survival probability to be simplified,
\bea
P(\gamma \to \gamma, \omega) & = & \prod_i \left\{ 1 - \epsilon_i \left( \frac{\omega}{\omega_0} \right)^2  \sin^2 \left[ \Delta_i \left( \frac{\omega_0}{\omega} \right) \right] \right\} \nn \\
& \simeq & 1 - \sum_i \epsilon_i \left( \frac{\omega}{\omega_0} \right)^2 \sin^2 \left[ \Delta_i \left( \frac{\omega_0}{\omega} \right) \right],
\eea
as by definition of the small conversion regime, any $\mc{O}(\epsilon^2)$  $\gamma \to a \to \gamma$ re-conversion terms are negligible.

This survival probability modulates the arriving photon spectrum. We shall assume
that we know the correct functional form of the original source spectrum of photons, which we
denote as $\mc{M}(\omega)$. For the purpose of this paper, we shall take this as a featureless power law plus an Fe K$\alpha$ line,
although for more complicated spectra, this may also need to include more emission and absorption lines
at (known) atomic energies.

The spectrum of (measured) arriving photons $\mc{A}(\omega)$ is then the source spectrum modulated by the
$P(\gamma \to \gamma, \omega)$ survival probability,
\bea
\mc{A}(\omega) & = & P(\gamma \to \gamma, \omega)\mc{M}(\omega) \nn \\
& = & \left\{ 1 - \sum_i \epsilon_i \left( \frac{\omega}{\omega_0} \right)^2 \sin^2 \left[ \Delta_i \left( \frac{\omega_0}{\omega} \right) \right] \right\} \mc{M}(\omega).
\eea
This can be decomposed into global and oscillatory terms:
\be
\mc{A}(\omega) = \left\{ 1 - \sum_i \frac{\epsilon_i}{2} \left( \frac{\omega}{\omega_0} \right)^2
+ \sum_i \frac{\epsilon_i}{2} \left( \frac{\omega}{\omega_0} \right)^2 \cos \left[ 2 \Delta_i \left( \frac{\omega_0}{\omega} \right) \right] \right\} \mc{M}(\omega)
\ee
The distinctive ALP-induced modulations lie in the sinusoidal terms, which average to zero.
To isolate these we want to first remove the global term, and so the function we should therefore fit to the data
is not $\mc{M}(\omega)$ itself, but instead
\bea
\label{globalred}
\mc{M}_a(\omega) & = & \left[ 1 - A \left( \frac{\omega}{\omega_0} \right)^2 \right] \mc{M}(\omega).
\eea
In the presence of axions, a fit to $\mc{M}_a(\omega)$ will produce oscillatory residuals
equally distributed about zero. These residuals will then take the form
\bea
\mc{R}(\omega) & = & \mc{A}_{true}(\omega) - \mc{A}_{fitted}(\omega) \nn \\
& = & \sum_i \frac{\epsilon_i}{2} \left( \frac{\omega}{\omega_0} \right)^2 \cos \left[ 2 \Delta_i \left( \frac{\omega_0}{\omega} \right) \right]
 \mc{M}(\omega),
\eea
where the prefactor of $\mc{M}(\omega)$ is a small quantity.
Taking the data/model ratio, we therefore obtain the fractional residual
\be
\mc{F}(\omega) = \frac{\mc{A}(\omega) - \mc{M}_a(\omega)}{\mc{M}(\omega)}  = \sum_i \frac{\epsilon_i}{2} \left( \frac{\omega}{\omega_0} \right)^2 \cos \left[ 2 \Delta_i \left( \frac{\omega_0}{\omega} \right) \right].
\ee
In this expression we have self-consistently neglected terms that are second-order in smallness - if such terms are important, the simple overall $\chi^2$ fit will have already revealed the failure of the `standard' astrophysical fit.

The above quantity is purely numerical and equally distributed about zero.
As expressed above, it is a function of energy. However, the $\omega^{-1}$ terms imply this
is not best analysed in
frequency space. Rather, by rewriting it in wavelength space
and multiplying through by $\lambda^2$, we obtain
\bea
\label{residcos}
\left( \frac{\lambda}{\lambda_0} \right)^2 \mc{F}(\lambda) & = & \sum_i \frac{\epsilon_i}{2} \cos \left[ 2 \Delta_i \left( \frac{\lambda}{\lambda_0} \right) \right].
\eea
We now have a quantity which can be analysed in a reasonably direct fashion as a Fourier series.
The contributing Fourier frequencies depend on the coherence lengths and electron densities of the individual domains.
So long as these remain broadly similar throughout the cluster,
by determining $\lambda^2 \mc{F}(\lambda)$ from data, we can search for axion-induced structure
by performing a Fourier decomposition of it.

\section{ALP search strategies}

The oscillatory structure of the residuals described in equation~\eqref{residcos} can be exploited using any of 
the following three strategies:
\begin{itemize}
 \item An analysis of the Fourier transformed data,
 \item A sinusoidal fit to the data,
 \item A machine learning approach. 
\end{itemize}
The Fourier transform of the dataset can be directly calculated using Fast-Fourier-Transform methods. 
As opposed to a fit, this does not require optimization of any of the parameters involved. Fitting to a small number of sinusoidal functions can be an efficient way to capture any oscillatory structure present in the residuals. The machine learning approach uses the complete list of residual data as an input to `learn' from a training set what value of $g$ is associated. Contrary to the Fourier and fit method, the machine learning method does not rely on the data being of oscillatory nature. It can be thought of as a universal fit: once the network is trained it includes all the relevant information that influence the conversion of photons to axions along the line of sight. Hence, it can be used as a benchmark for the Fourier and fit method: once those methods perform as well as the somewhat brute force machine learning approach, we know that we are able to resolve - and understand - all the possible structure of photon to ALP conversion imprinted onto the residuals.

In the following, we present the three methods in more detail, allowing a comparison between each of the three.

\subsection{{\it Analysis of Fourier-transformed Data}} \label{Fourier_sec}

Given a dataset analysed into the form of Eq. (\ref{residcos}),
we want to perform a Fourier analysis of the fractional residuals
\bea \label{uiyi}
 (u_i,y_i) & = & \left(\omega_i^{-1},\omega_i^{-2}\mc{F}(\omega_i)\right)\,,
\eea
where $\omega_i$ is the energy of the $i$-th bin. In general, we do not expect
the values $u_i$ to be equispaced.
We therefore require the tools of nonequispaced Fourier analysis and in particular the calculation of the nonequispaced discrete Fourier transform (NDFT). In this context, we make use of the tools developed in the context of nonequispaced fast Fourier transforms (NFFT).

The NDFT is defined\footnote{For example, see https://www-user.tu-chemnitz.de/~potts/nfft/} by expanding a function $y(x)$
specified on a total of $M$ data points
with respect to $N$ trigonometric polynomials as
\bea
 y(x) & = & \sum_{k=-\frac{N}{2}}^{\frac{N}{2}-1} \hat y_k\, e^{-2 \pi\,i\,k x}.
\eea
For a dataset of values $y_j$ and nonequispaced nodes $x_j$ with $j=0,..,M-1$ we can write
\bea
\label{ndft}
 y_j \equiv y(x_j) & = & \sum_{k=-\frac{N}{2}}^{\frac{N}{2}-1} \hat y_k\, e^{-2 \pi\,i\,k x_j},
\eea
which can be written as $\boldsymbol{y} = \boldsymbol{A}\, \boldsymbol{\hat y}$
with
\bea
 \boldsymbol{y} & = & (y_j)_{j=0,..,M-1}\, ,\\
 \boldsymbol{A} & = & (e^{-2 \pi\,i\,k x_j})_{k=-N/2,..,N/2-1; j=0,..,M-1}\,, \\
 \boldsymbol{\hat y} & = & (\hat y_j)_{j=0,..,M-1}\,.
\eea
The Fourier transform $\boldsymbol{\hat y}$ can hence be calculated from the inverse of the $N\times M$ matrix $\boldsymbol{A}$.

For periodicity the nodes $x_j$ are to be chosen in the interval $-\frac12 \leq x_j < \frac12$. Hence, we linearly rescale
the nodes of our dataset to place them in this interval,
\bea
 x_j & = & \frac{M-1}{M}\,\frac{u_j - u_0}{u_{M-1} - u_0}-\frac12 .
\eea
Note that for the case of equispaced nodes
\bea
 u_j & = & u_0 + \frac{j}{M-1}\,(u_{M-1}-u_0),
\eea
Equation (\ref{ndft}) reduces to the standard discrete Fourier transform
\bea
 y_j & = & \sum_{k=-\frac{N}{2}}^{\frac{N}{2}-1} \hat y_k\, e^{-2 \pi\,i\,k \left(\frac{j}{M} - \frac12 \right)},
\eea
for $j=0,..,M-1$.

If $N$ is large enough, this of course gives a perfect fit. However,
if photon-ALP conversion occurs then the dataset should be well described by a restricted number of
Fourier modes. By truncating this sum,
\bea
\label{Kmaxfourier}
 y^{(K)}(x) & \equiv & \text{Re}\left( \sum_{k=-K}^{K-1} \hat y_k\, e^{-2 \pi\,i\,k x} \right),
\eea
we obtain an approximation of the data to mode order $K \leq M/2$ (in the case of $K=M/2$, by Eq. (\ref{ndft}) $y^{(K)}(x)$ passes through all real data points $(x_i,y_i)$).

The essence of the axion search is that we expect the
axions to introduce modulations that allow this data to be well approximated by only its low frequency Fourier modes, i.e.~$K\ll M/2$ (at least in the limit of a very large number of datapoints).

\subsection{{\it Sinusoidal fit}} \label{Fit_sec}

For this method we simply fit the residuals \eqref{uiyi} that are expected to behave sinusoidally in the presence of ALPs. As fitting functions, we use $N_{\text{fit}}$ sine functions with amplitudes $A_i$ and phases $\phi_i$:
\bea \label{sinefit}
 y_{\text{fit}}^{(N_{\text{fit}})}(u) = \sum_{i=1}^{N_{\text{fit}}} A_i \sin\left( f_i\, u + \phi_i \right)\,.
\eea
In the spirit of a Fourier analysis, we fix the frequencies of the sine functions in \eqref{sinefit} to be
\bea
 f_i = \frac{2 \pi\,i}{u_{M}-u_0}\,,
\eea
i.e., the lowest frequency mode $i=1$ has one period in the interval $u_0 < u < u_M$ and the higher frequency modes have $i$ periods. The fit to the data then involves minimization of a $\chi^2$-function with $2 N_{\text{fit}}$ fit parameters. In principle, we could also treat the frequencies $f_i$ as fit parameters but in practice we find that the fit converges too slowly in this case and the amplitudes and phases allow enough flexibility of the fit function to capture the ALP induced oscillations.

\subsection{{\it Machine learning}} \label{ml_sec}

We set up a deep neural network (DNN) with Tensorflow~\footnote{We use the DNNRegressor estimator. For more details see the Tensorflow documentation.} that uses the list of residuals $\mc{F}_i$ in its entirety as `feature columns' while the value of $g$ that has been used to simulate those residuals is used as the `label column'. The design parameters of the DNN are the number of layers and the number of nodes as well as the loss function that the network tries to optimize. For the loss function we use the sum of squared errors between the actual and the DNN predicted values of the residuals $\mc{F}_i$. The data is divided into a training (75\% of data) and a test set (25\% of data). The training data is used to teach the network which value of $g$ is associated to a list of residuals by optimizing the parameters of each node while the test data is used to test how well the DNN performs and to estimate the prediction error. The number of layers and nodes are chosen such that the difference between the test error and training error is minimized, i.e. the network is build with enough flexibility to capture the data but avoids overfitting. We find this optimal value to be at 2 layers and 200 (40) nodes for Athena (XMM).

\section{ALP Data simulation}

We now aim to test this method on simulated data for XMM-Newton and Athena. Our focus here is on comparison of this method to a pure overall $\chi^2$ fit (and hence
whether it is possible to improve statistical sensitivity), rather than on applications to any particular real dataset or to precise
forecasts of future telescope sensitivity.

However, as we do want our simulated dataset to be motivated by real physics,
we take as a starting point for our simulations an absorbed power law together with an iron line at 6.4 keV,
reflecting a typical AGN spectrum.
We consider the simulated spectra of the model
\bea \label{fakeit}
 wabs * ALP * (zpowerlw + zgauss).
\eea{equation}
This represents the initial AGN spectrum, modulated by the effects of photon-ALP conversion, and then absorbed by the galactic column density of neutral Hydrogen. Here $ALP$ represents the effect of photon-ALP conversion within the cluster magnetic field.

As one of the most promising targets for ALP-photon conversion is the central Perseus AGN NGC1275,
we let $ALP$ represent simulated survival probabilities for the Perseus cluster environment, i.e. for ALP-photon conversion within an electron density $n_e$ and magnetic field model representing that of the Perseus cluster. As the B-field properties can only ever be known
statistically, and each explicit realization leads to different conversion probabilities even for statistically identical B-field parameters, we need to marginalize over many realizations of the magnetic field in Perseus. 

The properties of the magnetic field that enter the simulations are those used for Perseus in~\cite{160501043}: We use a central magnetic field strength of 25 $\mu$G~\cite{Taylor2006} and assume that $B$ decreases with radius as in the Coma cluster $B(r) \propto [n_e(r)]^{0.7}$~\cite{1002.0594}. The electron density as a function of radius in Perseus is~\cite{Churazov2003}
\bea
 n_e(r) = \frac{3.9 \times 10^{-2}}{\left[1+ \left(\frac{r}{80\,\text{kpc}} \right)^2 \right]^{1.8}} + \frac{4.05 \times 10^{-3}}{\left[1+ \left(\frac{r}{280\,\text{kpc}} \right)^2 \right]^{0.87}}\,\text{cm}^{-3}\,.
\eea
Finally, the domain length is a random parameter that is drawn from a powerlaw distribution with power 0.8 and with values between 3.5 and 10 kpc, motivated by~\cite{Vacca:2012up}. Each magnetic field simulation comprises a total of 300 domains. In the simulation, the mixed axion-photon state propagates through each magnetic field realization with the above parameters.

We simulate ALP to photon couplings in the range
\bea
 10^{-14} \text{GeV}^{-1} \leq g \leq 10^{-11} \text{GeV}^{-1}\,.
\eea
This represents a range from values beyond observational reach to those that have previously been excluded.

Based on XMM-Newton and predicted Athena responses\footnote{ These are taken from\\ http://hea-www.cfa.harvard.edu/~jzuhone/soxs/responses.html}, we generate fake data using the xspec command \texttt{fakeit}  with a simulated exposure time of 500 ks. \texttt{fakeit} simulates Poisson errors based on the exposure time. In terms of fitting, we work with an energy interval between 1.5 and 4.5 keV for XMM-Newton and 0.7 to 1.4 keV for Athena. This allows the demonstration of the ability of Athena (or any other telescope with microcalorimeter level resolution) to resolve the rapid ALP-induced oscillations that would be present at lower energies. For a real dataset, one would of course use the full energy range of the telescope.

We now describe the methods used to analyse the fake datasets:
\begin{enumerate}
 \item We first fit a zero model
 \bea
  wabs * ALPGlob * (zpowerlw + zgauss)
 \eea
 spectrum to the data. The $ALPGlob$ component represents the global modulation due to ALPs given in \eqref{globalred} and is given as
 \bea
  ALPGlob = 1- A\, \omega^2\,,
 \eea
 where $A = \omega_0^{-2} \sum \epsilon_i$. 
 
 In the absence of ALPs, one would expect the best-fit value of $A$ to vanish. In the presence of ALPs, while such a term removes the 
 `global' effects it will not account for the oscillations contained within the $ALP$ component in (\ref{fakeit}). 
 Hence, for a strong axion-photon coupling $\chi_0^2/$d.o.f would be significantly larger than one.
 \item We perform a Fourier analysis (as described in Section \ref{Fourier_sec}) and a sinusoidal fit (as described in Section \ref{Fit_sec}) of $(x_i,y_i) = \left(\omega_i^{-1},\omega_i^{-2}\mc{F}(\omega_i)\right)$  for a given $K$.\footnote{For Athena data we rebin two datapoints to one datapoint to speed up the calculation as half the energy resolution of Athena is easily sufficient to resolve ALP modulations.} 
   If low mode oscillations are present in the data, then  
 \bea
 \label{chi2stand}
  \chi_1^2 & = & \sum_{i=1}^N \frac{\left[ \mc{F}(\omega_i) - \omega_i^2\, y(\omega_i) \right]^2 }{\sigma_i^2}
 \eea
 per d.o.f will improve significantly compared to $\chi_0^2$. The fitfunction $y(\omega)$ equals $y^{(K)}(\omega_i) $ for the Fourier method and $y^{(N_{\text{fit}})}(\omega_i)$ for the sinusoidal fit.
 
The Fourier function $y^{(K)}(\omega)$ generally overestimates the fluctuations in the simulated data $\omega_i^{-2}\mc{F}(\omega_i)$ at the edges of the energy intervals. This is particularly significant for XMM data where the residuals are often constant for large $\omega \lesssim 4.5$ keV while $y^{(K)}(\omega)$ can oscillate strongly. This would lead to a bad fit to the data which is why we use a smoothing function $S(\omega)$ that multiplies $y^{(K)}(\omega)$ to smooth it out at the edges of the energy interval. For the Fourier method only, $\chi_1^2$ is then calculated as
\bea
 \label{chi2Fourier}
  \chi_{1,\text{Fourier}}^2 & = & \sum_{i=1}^N \frac{\left[ \mc{F}(\omega_i) - \omega_i^2\, S(\omega_i)\,y^{(K)}(\omega_i) \right]^2 }{\sigma_i^2}\,,
 \eea
 where we choose $S(\omega) = 1 \text{keV}^2/\omega^2$ as a smoothing function.
 \item The DNN is fitted by using 75\% of the residuals $\mc{F}_i$ as training data. For the remaining 25 \% of the residual data set the DNN predicts the value of $g$. 
\end{enumerate}
 
\begin{figure*}
\includegraphics[width=0.44\textwidth]{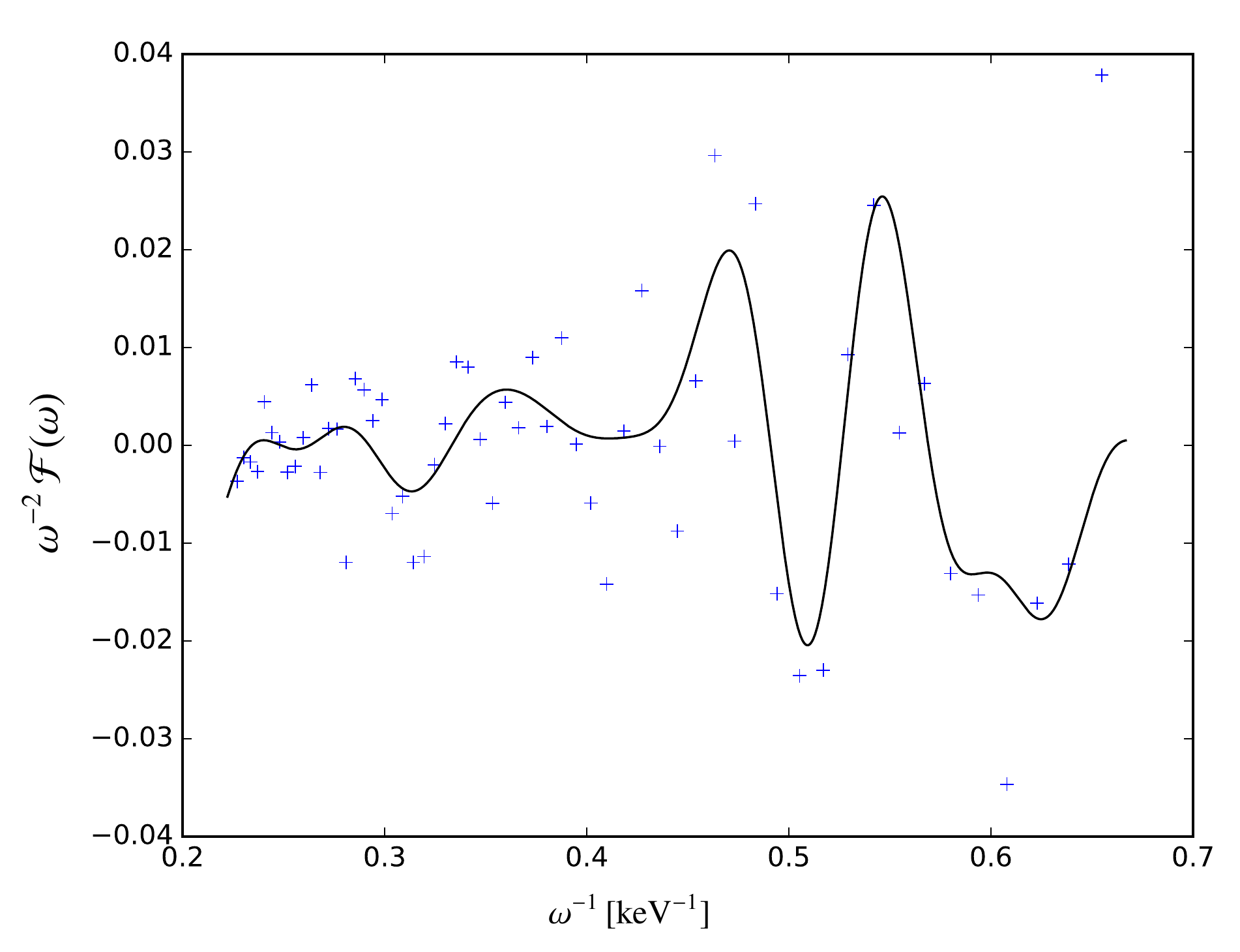} 
\includegraphics[width=0.44\textwidth]{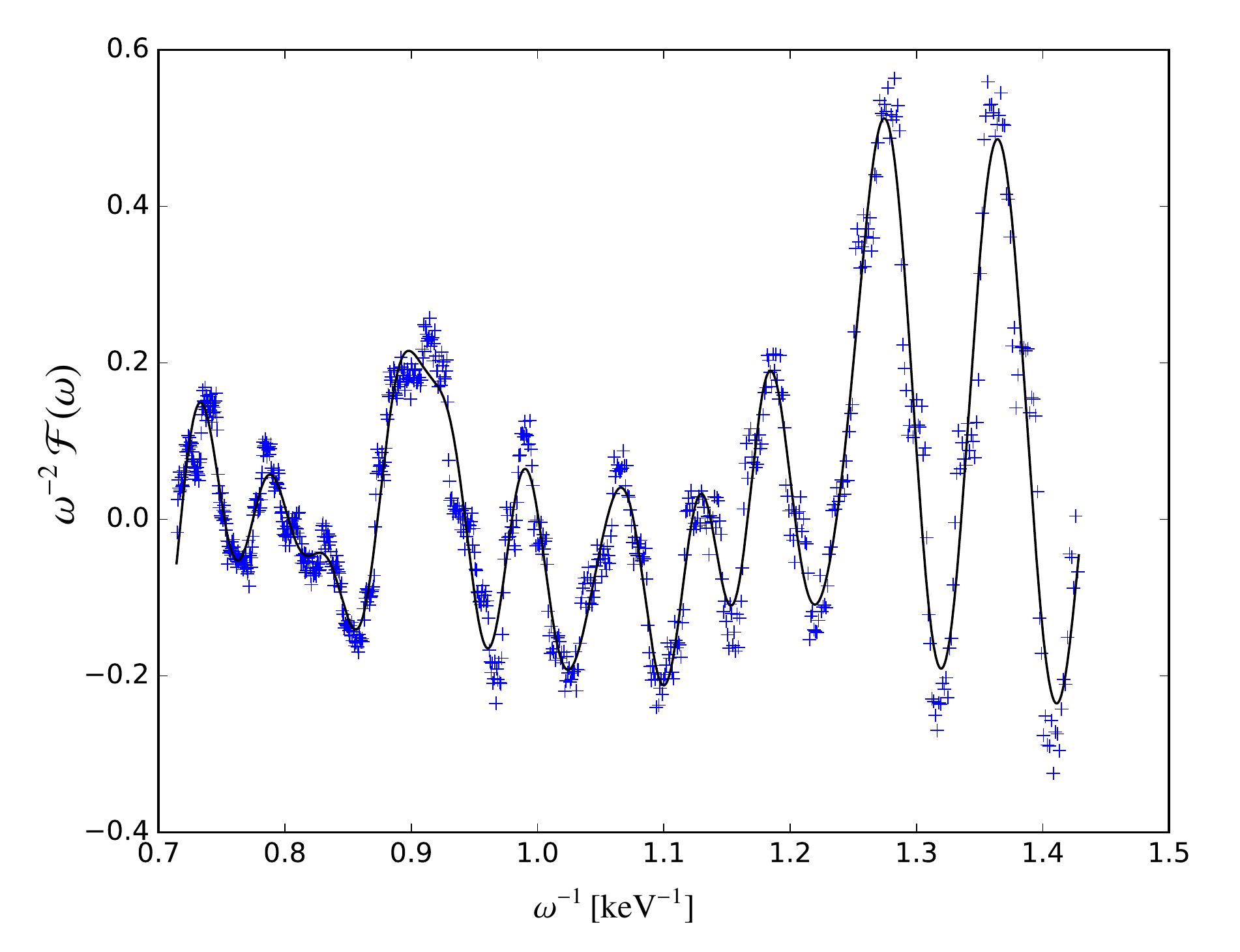}
\caption{For particular B-field and Poisson error realizations, we show plots of residuals $y_i = \omega_i^{-2}\mc{F}(\omega_i)$ vs low mode sinusoidal fitfunctions (left-hand $y_{\text{fit}}^{(8)}$  for XMM, right-hand $y^{(16)}$ for ATHENA) as a function of inverse energy. The ALP to photon coupling is $g = 5 \cdot 10^{-12} {\rm GeV}^{-1}$. }
\label{FourierExampleXMM_fig}
\end{figure*}

We show in Figure \ref{FourierExampleXMM_fig} how low Fourier modes can capture residuals in the data, for XMM-Newton and Athena respectively.
For the Fourier and fit method, whatever the precise method used, 
the quantity we are ultimately interested in is
\bea
 \Delta \chi^2 (g) & = & \chi_0^2 (g) - \chi_1^2 (g),
\eea
where $g$ is the ALP photon coupling. This represents the ability to improve the fit by including Fourier modes ($\chi_0^2(g)$ is equivalent to the
fit statistic with no Fourier modes included). The methods described in this paper are most relevant when $\chi_0^2(g)$ represents an acceptable fit -- but a fit which can be significantly improved by adding Fourier modes. If $\chi_0^2(g)$ already gives an unacceptable fit, then
the techniques of this paper are largely superfluous.

Photon to ALP modulations are very well represented by the leading Fourier modes and hence this analysis is very sensitive to the presence of ALP physics. An atomic line for instance would lead to virtually no improvement in $\Delta \chi^2 (g)$ when using a Fourier analysis and hence would not be picked up by our method. The method is also very efficient as the calculation of Fourier coefficients can be performed very quickly and no spectral fit beyond the standard absorbed powerlaw is necessary.

Note that $\Delta \chi^2 (g=0)$ represents improvements in the fit due simply to Poisson noise - even in the absence of ALPs, it is expected that some improvement can be attained by fitting the Poisson noise of the residuals with sinusoidal functions. Constraints will be obtained when $\Delta \chi^2 (g) \gg \Delta \chi^2 (g=0)$.

For every $g$ we generate 40 realizations of the magnetic field in Perseus and subsequently 50 fake data sets with different Poisson errors. Hence, for every $g$ there are 2000 sets of simulated data.

\begin{figure}
\includegraphics[width=0.44\textwidth]{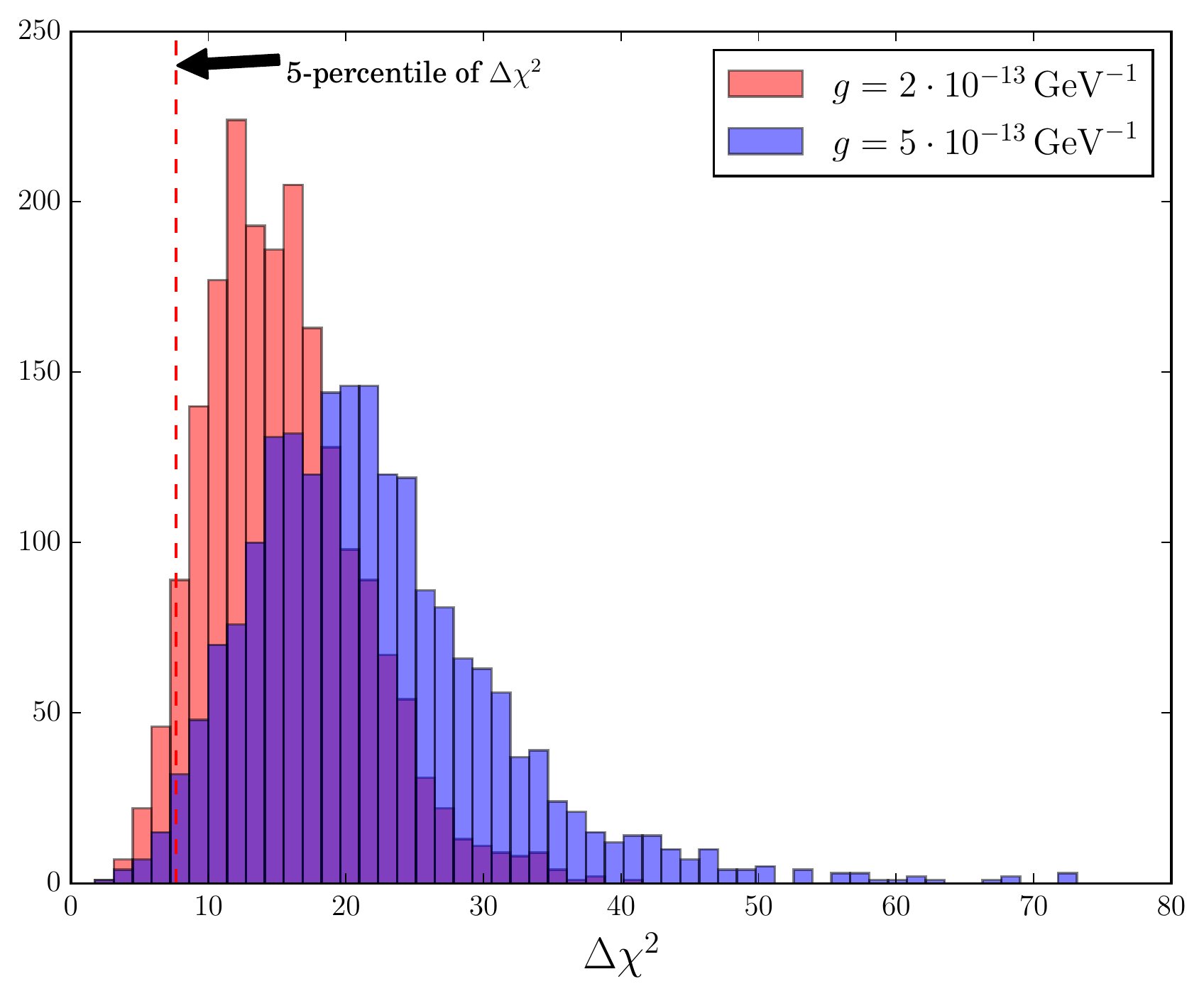}
\caption{Histogram of the $\Delta \chi^2$ distributions for $g = 2 \cdot 10^{-13} {\rm GeV}^{-1}$ (red) and $g = 5 \cdot 10^{-13} {\rm GeV}^{-1}$ (blue). As $g$ increases the distribution moves towards larger $\Delta \chi^2$. The 5th percentile of the simulated data is used to infer a bound on $g$, see main text. This plot uses the sinusoidal fit method for XMM-Newton simulated data.}
\label{95limexplained_fig}
\end{figure}

In order to set bounds on $g$ one has to compare the simulated $\Delta \chi^2$-distributions to $\Delta \chi^2_{\text{obs}}$ of the observed data. If $\Delta \chi^2_{\text{obs}}$ is lower than the 95th percentile of the $\Delta \chi^2 (g=0)$-distribution we cannot exclude that the modulations are purely due to Poisson noise and we cannot set any bounds. Let us now discuss the case that the observed $\Delta \chi^2$ is larger. We take as the null hypothesis the statement that ALPs exist with a given coupling $g$. The null hypothesis is excluded at 95\%, if 95\% of the simulated values of $\Delta \chi^2 (g)$ are larger than the measured $\Delta \chi^2_{\text{observed}}$. Since 
any percentile value of the $\Delta \chi^2 (g\neq0)$-distribution is expected to grow monotonically in $g$ the bound $g_{\text{bound}}$ is defined via
\bea
 \text{5th percentile}\, \left[ \Delta \chi^2(g_{\text{bound}}) \right]= \Delta \chi^2_{\text{obs}}.
\eea
If there large modulations are present in the data (i.e.~$\Delta \chi^2_{\text{observed}}$ is large) we can only set a weak bound on $g$, i.e.~$g_{\text{bound}}$ is large. However, if $\Delta \chi^2_{\text{obs}}$ is small we can set a stronger bound on $g$ and $g_{\text{bound}}$ is small.  These effects are illustrated in Figure~\ref{95limexplained_fig}, where we see that a larger coupling leads (statistically) to a larger $\Delta \chi^2(g)$.

For the DNN we can use similar criteria for exclusion when directly using percentages of the DNN predicted value of $g = G_\text{DNN}$ instead of $\Delta \chi^2$ as for the Fourier/fit methods. For real data, the set of measured residuals is fed into the DNN that has been trained with simulated data which results in a predicted value of $G_\text{obs}:=G_\text{DNN}(\mc{F}_{i,\text{obs}})$. If the DNN predicted value of $G_\text{obs}$ is lower than the 95th percentile of the values the DNN predicts for $g=0$ from the test set, then the residuals being purely due to Poisson noise cannot be excluded. Similarly to the previous discussion of setting bounds on $g$ for the Fourier and fit method, the 95 \% exclusion bound in the DNN method is define via
\bea
 \text{5th percentile}\, \left[ G_\text{DNN}(\mc{F}_{i,\text{sim}}\,|\,g_{\text{bound}}) \right]= G_{\text{obs}}\,,
\eea
where $G_\text{DNN}(\mc{F}_{i,\text{sim}}\,|\,g)$ are the DNN predictions on the simulated data for a given $g$ in the test set.

\section{Results}

In the following, we compare the standard $\chi^2_0$ calculation which was previously used to look for ALPs to the Fourier method explained in Section~\ref{Fourier_sec}, the sinusoidal fit method explained in Section~\ref{Fit_sec}, and the machine learning approach discussed in Section~\ref{ml_sec}. We apply these methods to the simulated data for XMM and Athena.

\subsection{{\it Athena}}

\begin{figure*}
\includegraphics[width=0.44\textwidth]{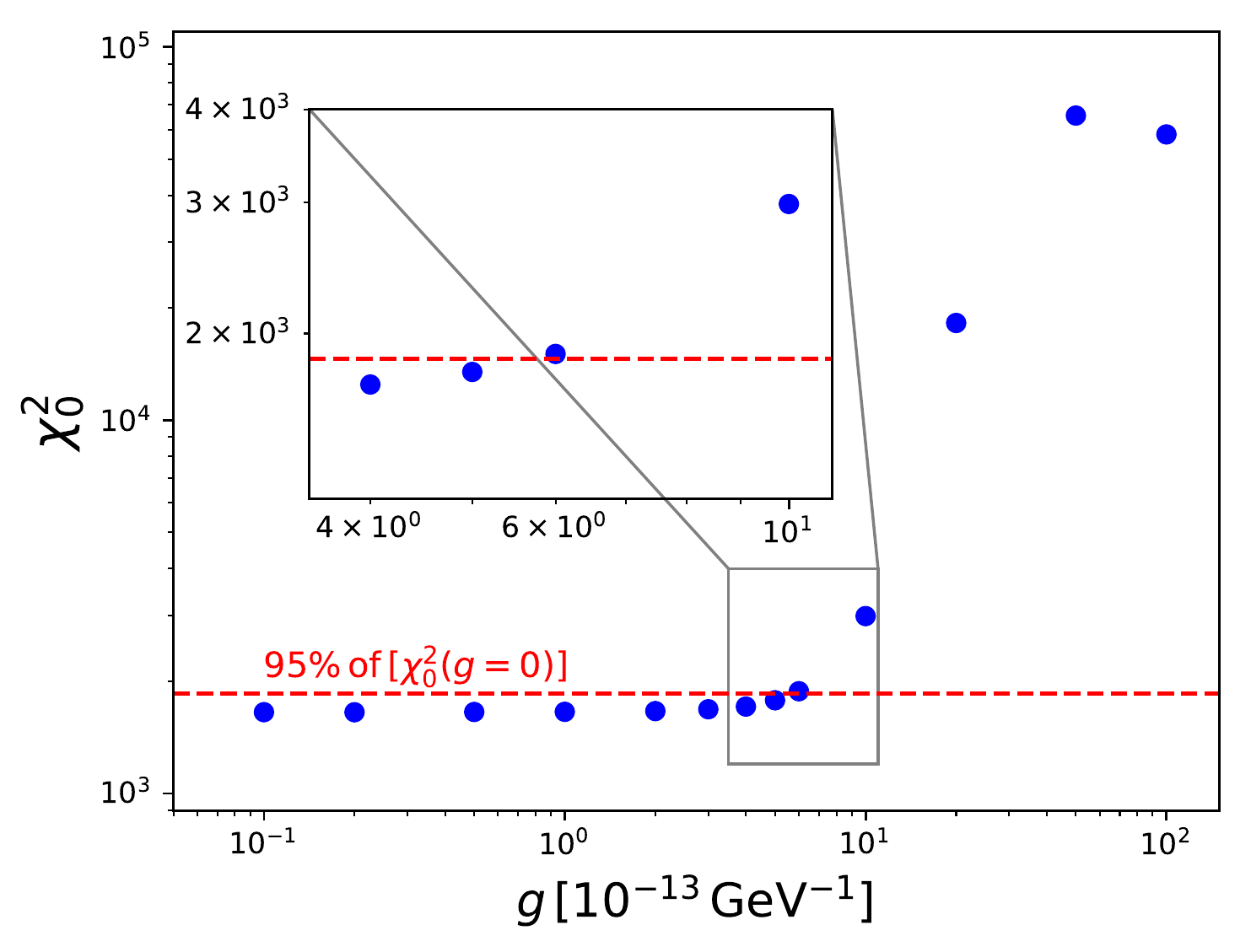} 
\includegraphics[width=0.44\textwidth]{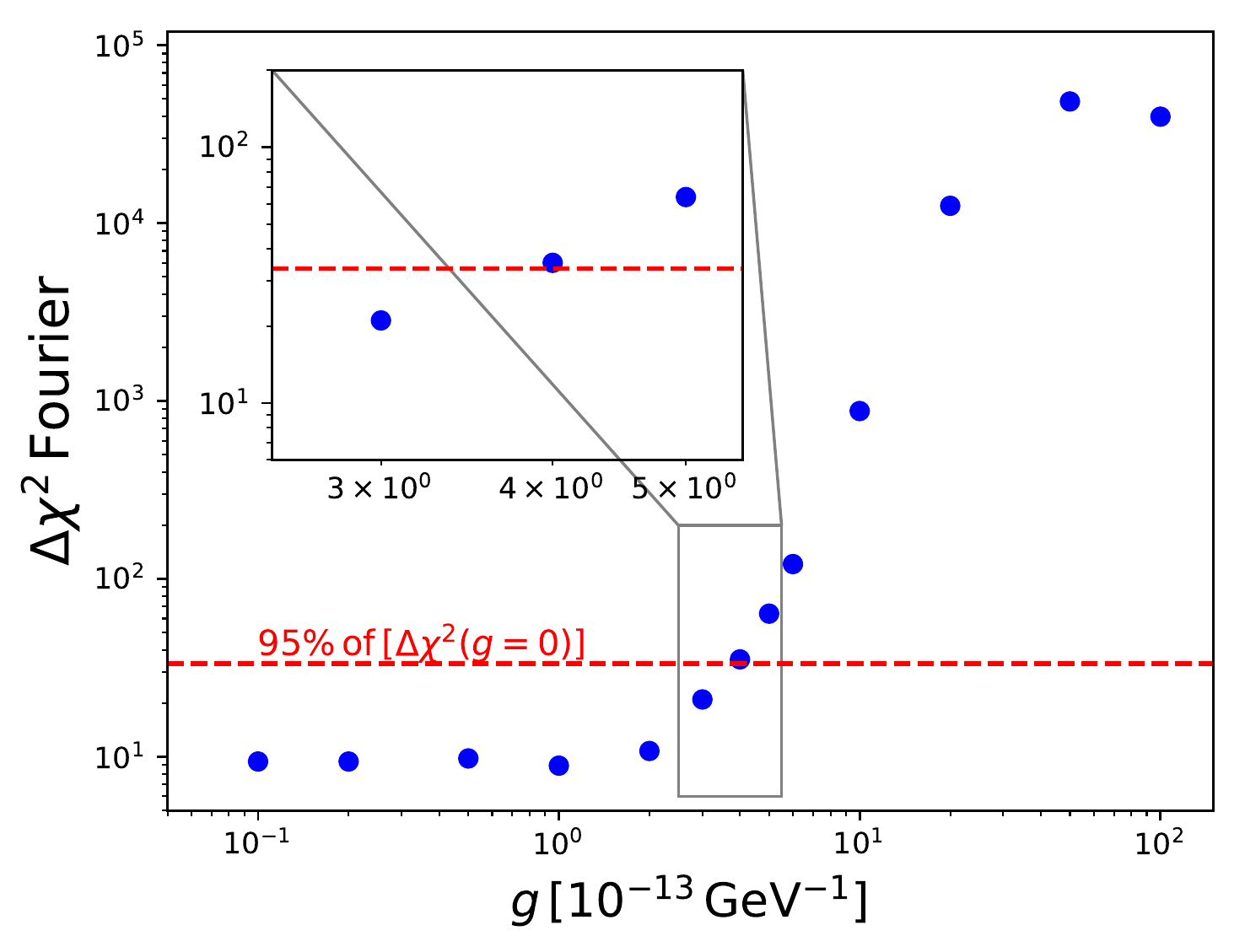}\\
\includegraphics[width=0.44\textwidth]{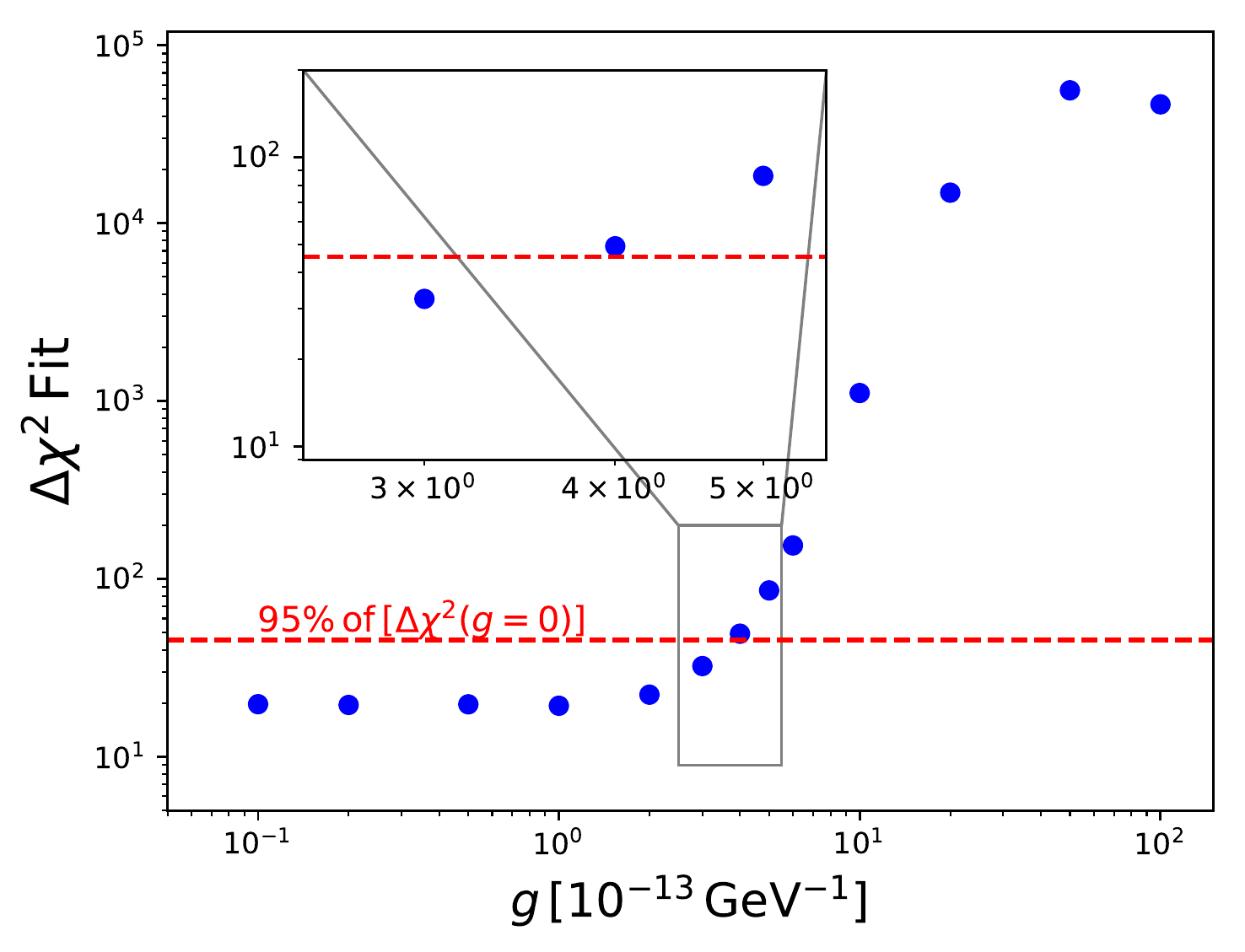}
\includegraphics[width=0.44\textwidth]{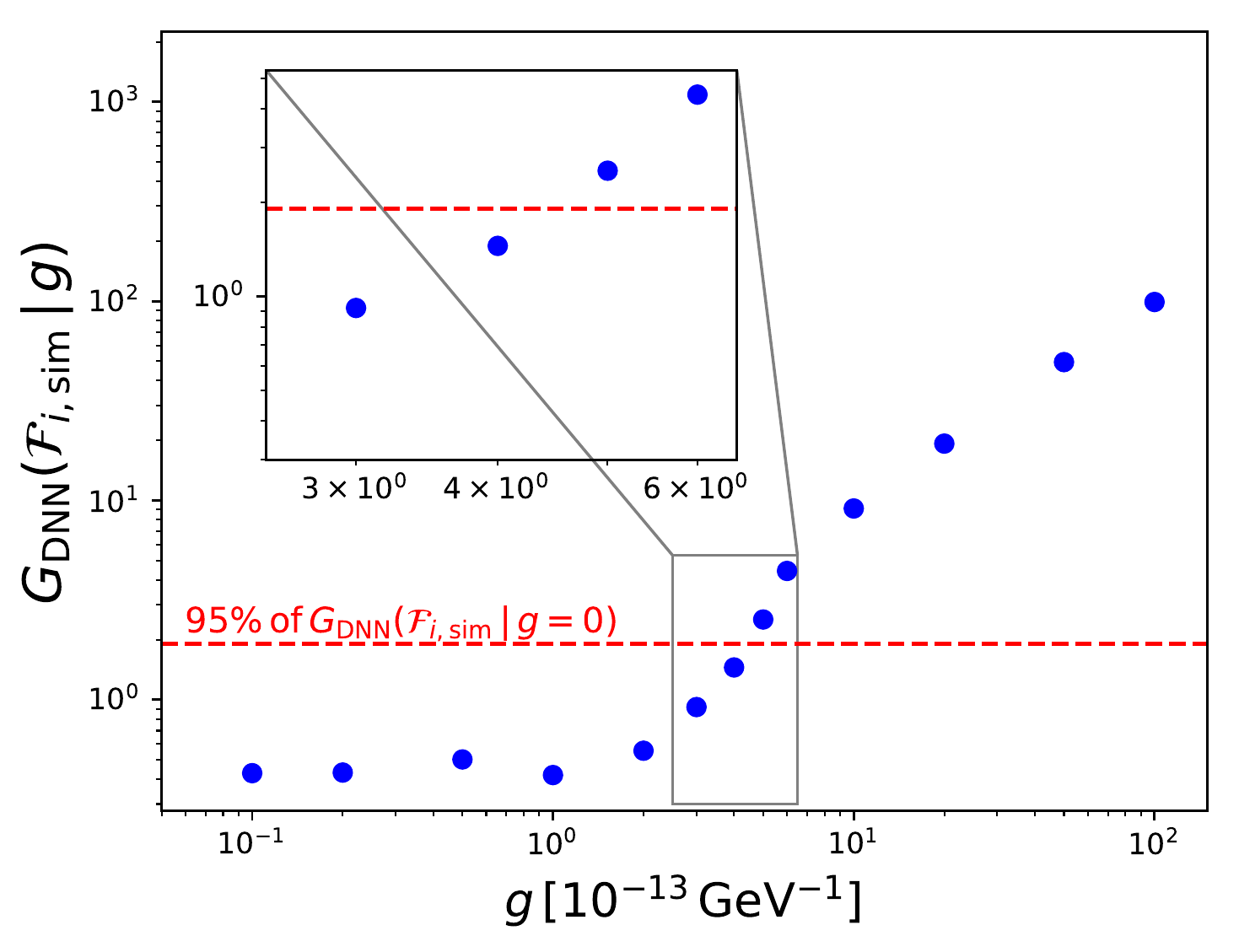}
\caption{The four results for Athena. From left to right, top to bottom, these plots show (a) 5-percentile of the powerlaw fit $\chi_0^2(g)$ for Athena. There are 1747 degrees of freedom for this fit. The dashed red line is the 95th percentile of $\chi^2_0 (g=0)$ and marks which values of $g$ can be excluded: a point above it can be excluded at 95\% confidence. (b) 5-percentile of $\Delta \chi^2(g)$ for the Fourier method with $\omega^{-2}$-weighting (see text) and $K = 16$ for Athena. (c) 5-percentile of $\Delta \chi^2(g)$ for a sinusoidal fit with $N_{\text{fit}} = 16$ for Athena. (d) 5th percentile of $G_\text{DNN}(\mc{F}_{i,\text{sim}}\,|\,g)$ for the DNN method for Athena. The dashed red line is the 95th percentile of $G_\text{DNN}(\mc{F}_{i,\text{sim}}\,|\,g=0)$ and marks which values of $g$ can be excluded: a point above it can be excluded at 95\% confidence.}
\label{Athenacombined_fig}
\end{figure*}

The standard powerlaw $\chi^2_0$ method, the weighted Fourier method, the sinusoidal fit method, and the DNN machine learning method are shown in Figure~\ref{Athenacombined_fig}. Every dataset contains $\sim$1800 points so for small $g< 10^{-13}$ GeV$^{-1}$ where the powerlaw is already a decent fit to the data, we expect a $\chi_0^2 \sim 1800$ which is what we find. Once we rebin two datapoints to one for the Fourier and fit method, the degrees of freedom are roughly divided by two so $\chi_0^2(\text{Fourier,Fit}) \sim 900$. $\chi_1^2$ is calculated once the Fourier modes or the sinusoidal fit function are included. Even for very small $g$ we expect these methods to slightly improve the fit, hence $\chi_1^2 < \chi_0^2$, effectively fitting Poisson noise. We find $\chi_1^2 \lesssim \chi_0^2$ for very small $g < 10^{-13}$ GeV$^{-1}$ as expected.  

As illustrated in these figures, the $\chi^2_0$ method gives a bound of $g \lesssim 5.9 \cdot 10^{-13}$ GeV$^{-1}$. The weighted Fourier method and the fit method do significantly better with a bound of $g \lesssim 3.8 \cdot 10^{-13}$ GeV$^{-1}$. The DNN method also does better giving a bound of $g \lesssim 4.2 \cdot 10^{-13}$ GeV$^{-1}$. This represents an improvement by a factor of 1.5 in the coupling compared to the $\chi^2_0$ method. As all actual physical effects (interconversion of photons and axions) scale as $g_{a \gamma \gamma}^2$, the improvement in sensitivity is then slightly greater than a factor of two.

\subsection{{\it XMM-Newton}}

\begin{figure*}
\includegraphics[width=0.42\textwidth]{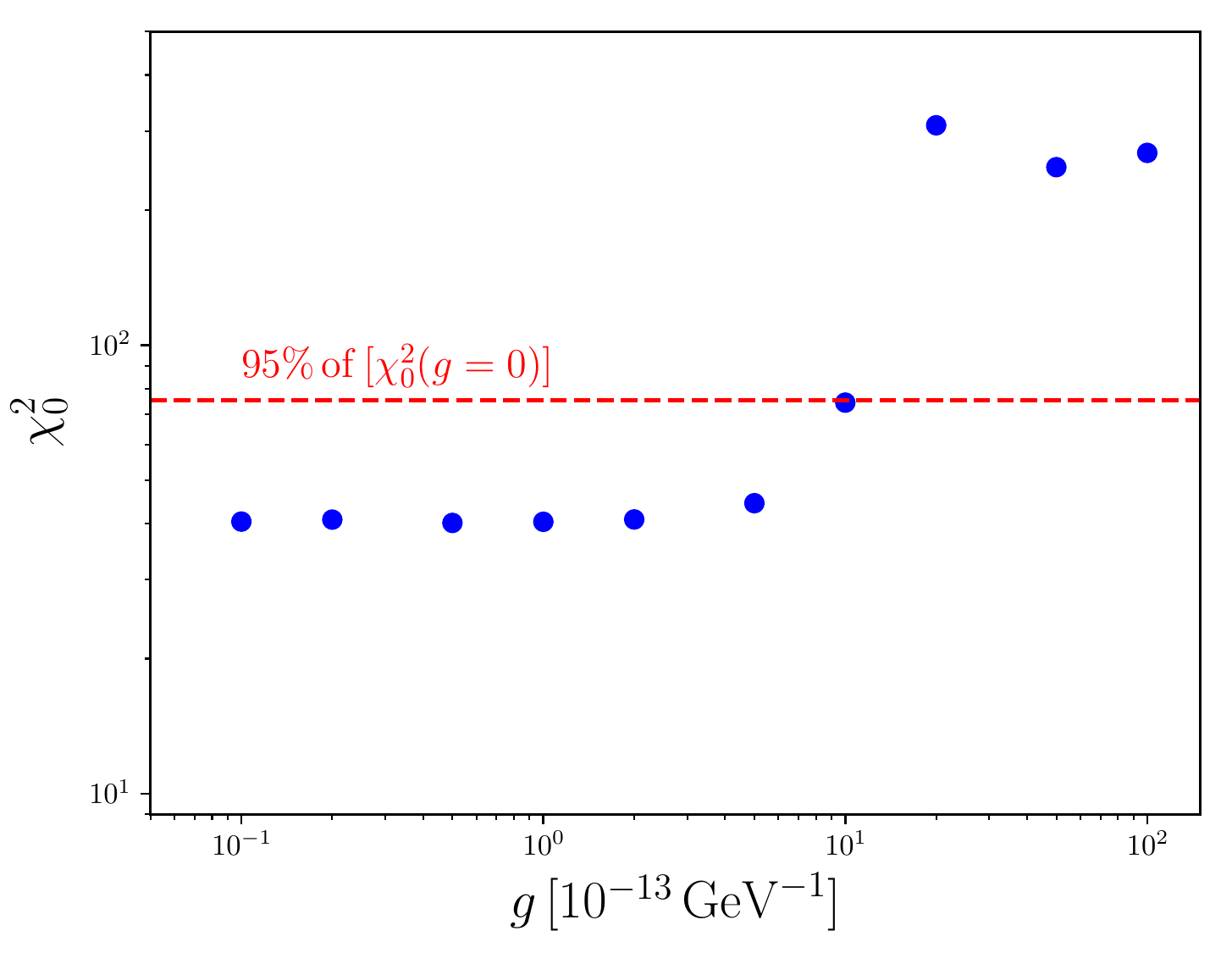} 
\includegraphics[width=0.42\textwidth]{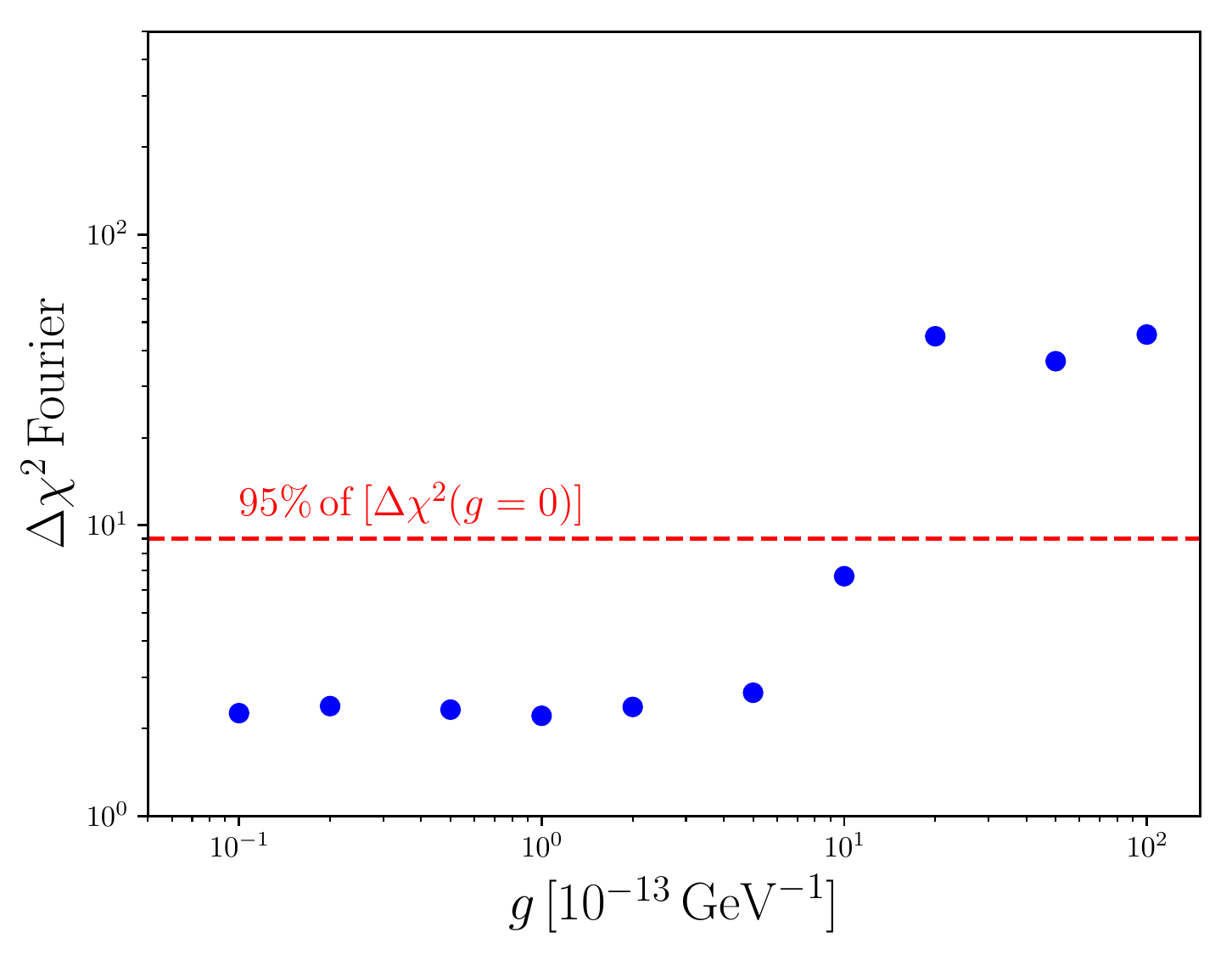} \\
\includegraphics[width=0.42\textwidth]{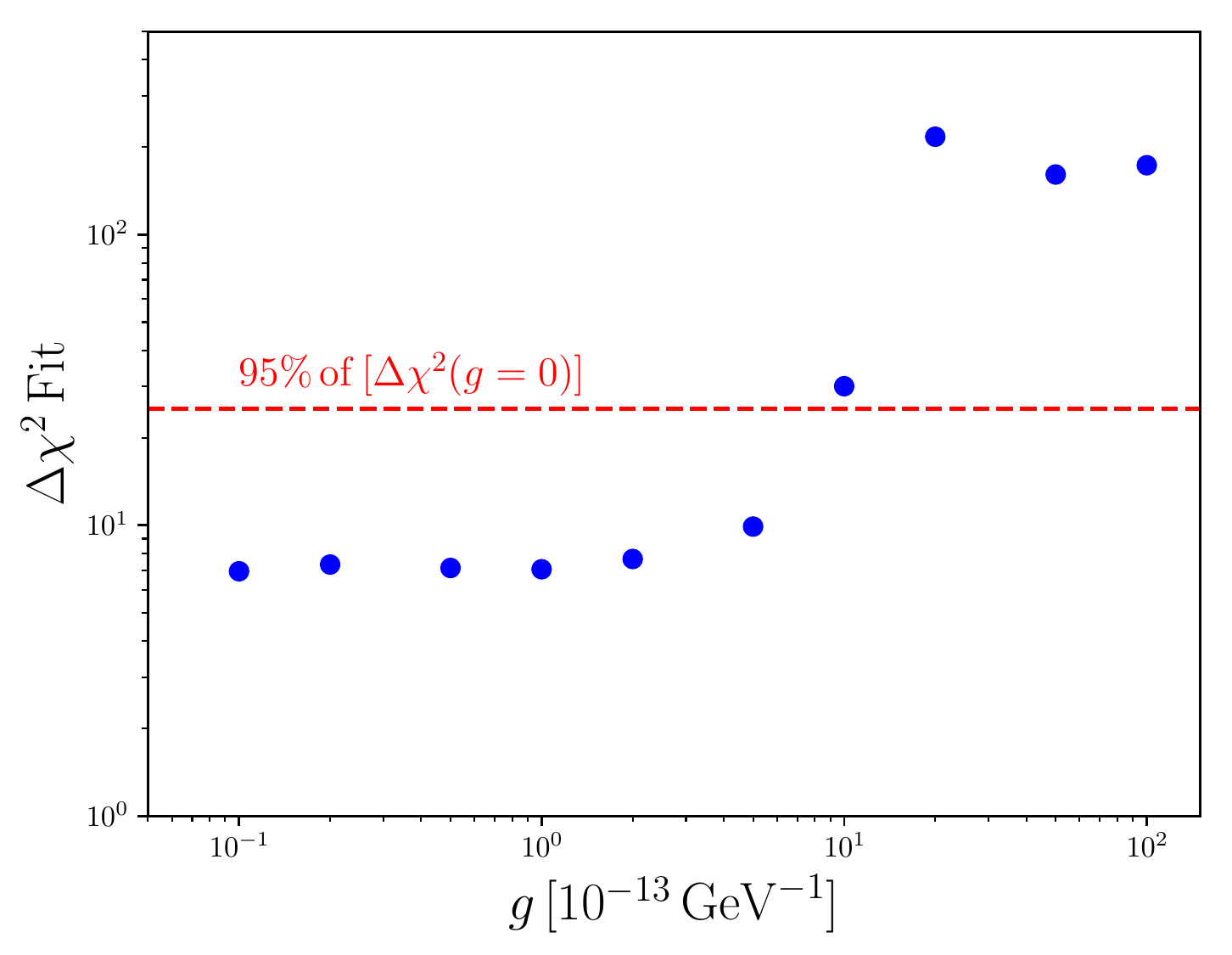} 
\includegraphics[width=0.42\textwidth]{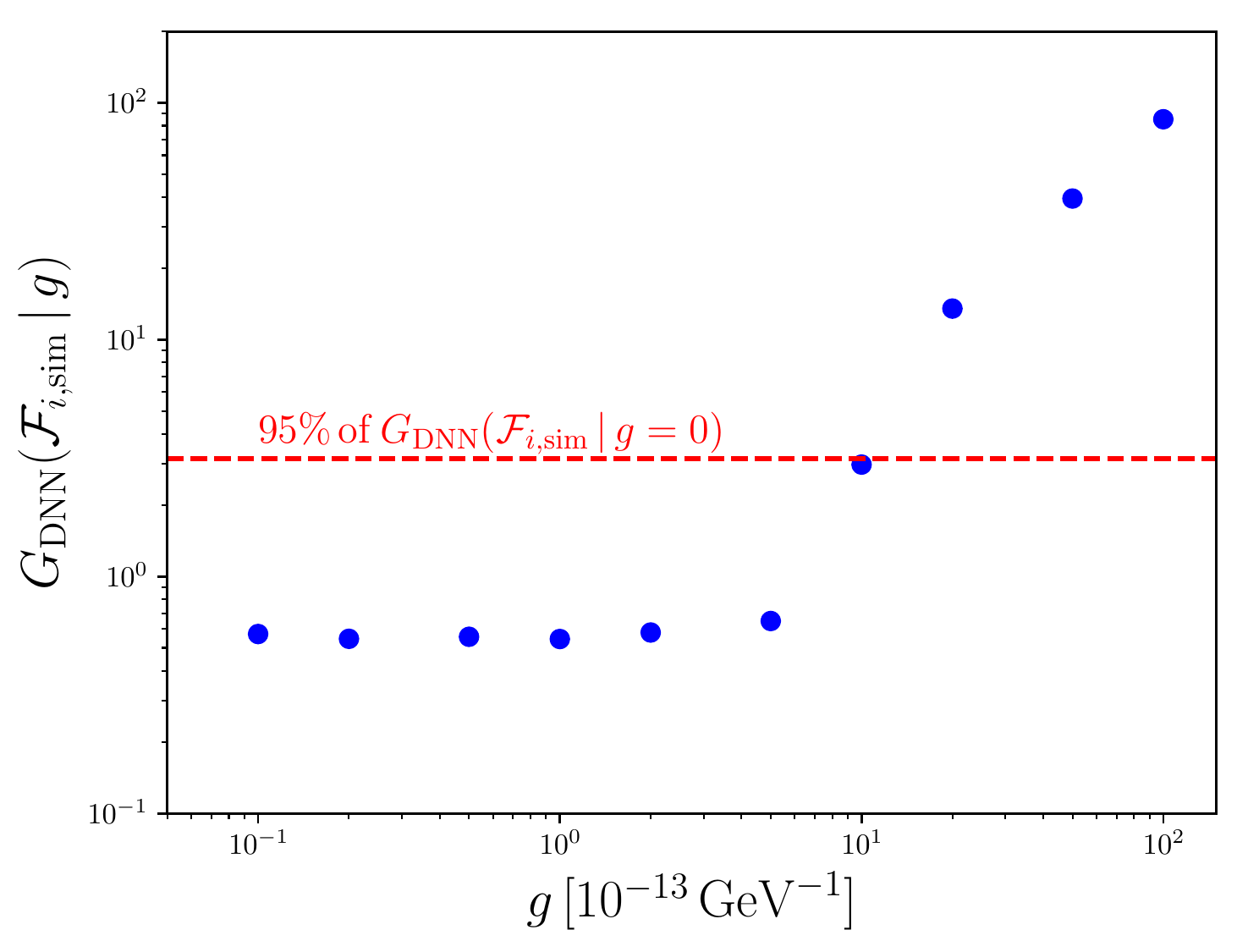} 
{\caption{The four results for XMM. From left to right, top to bottom, these plots show (a) 5th percentile of the powerlaw fit $\chi_0^2(g)$ for XMM. There are 56 degrees of freedom for this fit. The dashed red line is the 95th percentile of $\chi^2_0 (g=0)$ and marks which values of $g$ can be excluded: a point above it can be excluded at 95\% confidence. 
(b) 5th percentile of $\Delta \chi^2(g)$ for the Fourier method with $\omega^{-2}$-weighting (see text) and $K = 8$ for XMM. (c) 5th percentile of $\Delta \chi^2(g)$ for a sinusoidal fit with $N_{\text{fit}} = 8$ for XMM. 
(d) 5th percentile of $G_\text{DNN}(\mc{F}_{i,\text{sim}}\,|\,g)$ for the DNN method for XMM. The dashed red line is the 95th percentile of $G_\text{DNN}(\mc{F}_{i,\text{sim}}\,|\,g=0)$ and marks which values of $g$ can be excluded: a point above it can be excluded at 95\% confidence.}\label{XMMResults}}
\end{figure*}

The results for XMM-Newton are all shown in Figure~\ref{XMMResults}.
In this case, the sinusoidal fit method can exclude $g$ between $5 \cdot 10^{-13}$ GeV$^{-1}$ and $10^{-12}$ GeV$^{-1}$ and hence does slightly better than the $\chi^2_0$, weighted Fourier and DNN methods which can only exclude $g > 10^{-12}$ GeV$^{-1}$.

\section{Discussion}

X-ray observations of AGNs offer one of the most sensitive probes of axion-like particles in the low mass regime and it is therefore important to ensure that maximal statistical sensitivity is obtained in searches for them. In this paper we have developed analysis methods aimed at exploiting the characteristic quasi-sinusoidal modulations that are induced by ALPs. Based on simulated data, these methods appear to improve sensitivity to photon-axion conversion by a factor of two (in conversion probability).

\section*{Acknowledgements}

We thank Cliff Burgess, Francesca Day, Nicholas Jennings, Sven Krippendorf, David Marsh for discussions. This work was funded in part by ERC Starting Grant 307605. 
\bibliographystyle{hapj}
\bibliography{AxionStatsBib}

\begin{thebibliography}{32}
\expandafter\ifx\csname natexlab\endcsname\relax\def\natexlab#1{#1}\fi

\bibitem[{Agrawal {et~al.}(2018)Agrawal, Fan, Reece, \& Wang}]{Agrawal}
Agrawal, P., Fan, J., Reece, M., \& Wang, L.-T. 2018, JHEP, 02, 006, 1709.06085

\bibitem[{Ajello {et~al.}(2016)}]{Ajello}
Ajello, M., {et~al.} 2016, Phys. Rev. Lett., 116, 161101, 1603.06978

\bibitem[{Angus {et~al.}(2014)Angus, Conlon, Marsh, Powell, \&
  Witkowski}]{Angus:2013sua}
Angus, S., Conlon, J.~P., Marsh, M. C.~D., Powell, A.~J., \& Witkowski, L.~T.
  2014, JCAP, 1409, 026, 1312.3947

\bibitem[{Arvanitaki {et~al.}(2010)Arvanitaki, Dimopoulos, Dubovsky, Kaloper,
  \& March-Russell}]{Arvanitaki:2009fg}
Arvanitaki, A., Dimopoulos, S., Dubovsky, S., Kaloper, N., \& March-Russell, J.
  2010, Phys. Rev., D81, 123530, 0905.4720

\bibitem[{Berg {et~al.}(2017)Berg, Conlon, Day, Jennings, Krippendorf, Powell,
  \& Rummel}]{160501043}
Berg, M., Conlon, J.~P., Day, F., Jennings, N., Krippendorf, S., Powell, A.~J.,
  \& Rummel, M. 2017, Astrophys. J., 847, 101, 1605.01043

\bibitem[{Bonafede {et~al.}(2010)Bonafede, Feretti, Murgia, Govoni, Giovannini,
  Dallacasa, Dolag, \& Taylor}]{1002.0594}
Bonafede, A., Feretti, L., Murgia, M., Govoni, F., Giovannini, G., Dallacasa,
  D., Dolag, K., \& Taylor, G.~B. 2010, Astron. Astrophys., 513, A30, 1002.0594

\bibitem[{Burrage {et~al.}(2009)Burrage, Davis, \& Shaw}]{Burrage:2009mj}
Burrage, C., Davis, A.-C., \& Shaw, D.~J. 2009, Phys. Rev. Lett., 102, 201101,
  0902.2320

\bibitem[{Chen \& Conlon(2017)}]{Chen:2017mjf}
Chen, L., \& Conlon, J.~P. 2017, 1712.08313

\bibitem[{Churazov {et~al.}(2003)Churazov, Forman, Jones, \&
  Bohringer}]{Churazov2003}
Churazov, E., Forman, W., Jones, C., \& Bohringer, H. 2003, Astrophys. J., 590,
  225, astro-ph/0301482

\bibitem[{Cicoli {et~al.}(2012)Cicoli, Goodsell, \& Ringwald}]{Cicoli:2012sz}
Cicoli, M., Goodsell, M., \& Ringwald, A. 2012, JHEP, 10, 146, 1206.0819

\bibitem[{Conlon(2006)}]{Conlon:2006tq}
Conlon, J.~P. 2006, JHEP, 05, 078, hep-th/0602233

\bibitem[{Conlon {et~al.}(2018)Conlon, Day, Jennings, Krippendorf, \&
  Muia}]{Conlon:2017ofb}
Conlon, J.~P., Day, F., Jennings, N., Krippendorf, S., \& Muia, F. 2018, Mon.
  Not. Roy. Astron. Soc., 473, 4932, 1707.00176

\bibitem[{Conlon {et~al.}(2017)Conlon, Day, Jennings, Krippendorf, \&
  Rummel}]{1704.05256}
Conlon, J.~P., Day, F., Jennings, N., Krippendorf, S., \& Rummel, M. 2017,
  JCAP, 1707, 005, 1704.05256

\bibitem[{Conlon \& Marsh(2013)}]{Conlon:2013txa}
Conlon, J.~P., \& Marsh, M. C.~D. 2013, Phys. Rev. Lett., 111, 151301,
  1305.3603

\bibitem[{Conlon {et~al.}(2016)Conlon, Marsh, \& Powell}]{Conlon:2015uwa}
Conlon, J.~P., Marsh, M. C.~D., \& Powell, A.~J. 2016, Phys. Rev., D93, 123526,
  1509.06748

\bibitem[{Farina {et~al.}(2017)Farina, Pappadopulo, Rompineve, \&
  Tesi}]{Farina}
Farina, M., Pappadopulo, D., Rompineve, F., \& Tesi, A. 2017, JHEP, 01, 095,
  1611.09855

\bibitem[{Majumdar {et~al.}(2018)Majumdar, Calore, \& Horns}]{MajumdarHorns}
Majumdar, J., Calore, F., \& Horns, D. 2018, JCAP, 1804, 048, 1801.08813

\bibitem[{Marsh(2016)}]{Marsh}
Marsh, D. J.~E. 2016, Phys. Rept., 643, 1, 1510.07633

\bibitem[{Marsh {et~al.}(2017)Marsh, Russell, Fabian, McNamara, Nulsen, \&
  Reynolds}]{Marsh:2017yvc}
Marsh, M. C.~D., Russell, H.~R., Fabian, A.~C., McNamara, B.~P., Nulsen, P., \&
  Reynolds, C.~S. 2017, JCAP, 1712, 036, 1703.07354

\bibitem[{Payez {et~al.}(2012)Payez, Cudell, \& Hutsemekers}]{Payez}
Payez, A., Cudell, J.~R., \& Hutsemekers, D. 2012, JCAP, 1207, 041, 1204.6187

\bibitem[{Peccei \& Quinn(1977)}]{PecceiQuinn}
Peccei, R.~D., \& Quinn, H.~R. 1977, Phys. Rev. Lett., 38, 1440

\bibitem[{Powell(2015)}]{Powell:2014mda}
Powell, A.~J. 2015, JCAP, 1509, 017, 1411.4172

\bibitem[{Raffelt \& Stodolsky(1988)}]{RS}
Raffelt, G., \& Stodolsky, L. 1988, Phys. Rev., D37, 1237

\bibitem[{Schlederer \& Sigl(2016)}]{Schlederer:2015jwa}
Schlederer, M., \& Sigl, G. 2016, JCAP, 1601, 038, 1507.02855

\bibitem[{Sikivie(1983)}]{Sikivie:1983ip}
Sikivie, P. 1983, Phys. Rev. Lett., 51, 1415, [Erratum: Phys. Rev.
  Lett.52,695(1984)]

\bibitem[{Svrcek \& Witten(2006)}]{Svrcek:2006yi}
Svrcek, P., \& Witten, E. 2006, JHEP, 06, 051, hep-th/0605206

\bibitem[{Taylor {et~al.}(2006)Taylor, Gugliucci, Fabian, Sanders, Gentile, \&
  Allen}]{Taylor2006}
Taylor, G.~B., Gugliucci, N.~E., Fabian, A.~C., Sanders, J.~S., Gentile, G., \&
  Allen, S.~W. 2006, Mon. Not. Roy. Astron. Soc., 368, 1500, astro-ph/0602622

\bibitem[{Vacca {et~al.}(2012)Vacca, Murgia, Govoni, Feretti, Giovannini,
  Perley, \& Taylor}]{Vacca:2012up}
Vacca, V., Murgia, M., Govoni, F., Feretti, L., Giovannini, G., Perley, R.~A.,
  \& Taylor, G.~B. 2012, Astron. Astrophys., 540, A38, 1201.4119

\bibitem[{Weinberg(1978)}]{Weinberg}
Weinberg, S. 1978, Phys. Rev. Lett., 40, 223

\bibitem[{Wilczek(1978)}]{Wilczek}
Wilczek, F. 1978, Phys. Rev. Lett., 40, 279

\bibitem[{Wouters \& Brun(2012)}]{Wouters:2012qd}
Wouters, D., \& Brun, P. 2012, Phys. Rev., D86, 043005, 1205.6428

\bibitem[{Wouters \& Brun(2013)}]{Wouters:2013hua}
------. 2013, Astrophys. J., 772, 44, 1304.0989

\end{thebibliography}

\label{lastpage}
\end{document}